\documentclass[runningheads,a4paper]{llncs}

\usepackage{amssymb}
\usepackage{graphicx}

\usepackage{url}
\urldef{\mailsa}\path| mahyunst@yahoo.com, mahyuddin@usu.ac.id |
\newcommand{\keywords}[1]{\par\addvspace\baselineskip
\noindent\keywordname\enspace\ignorespaces#1}
\newcommand{\kotak}{\rule{.08in}{.08in}}
\newcommand{\kotakt}{\rule{.08in}{.12in}}
\begin{document}

\mainmatter  
\title{Simple Search Engine Model: Adaptive Properties for Doubleton}

\titlerunning{Simple Search Engine Model: Adaptive Properties for Doubleton}
\author{Mahyuddin K. M. Nasution
\authorrunning{Mahyuddin K. M. Nasution}
\institute{Information Technology Department, \\ Fakultas Ilmu Komputer dan Teknologi Informasi\\
Universitas Sumatera Utara, Padang Bulan, Medan 20155, Sumatera Utara, Indonesia\\
\mailsa\\}}
\toctitle{}
\tocauthor{}
\maketitle

\begin{abstract}
In this paper we study the relationship between query and search engine by exploring the adaptive properties for doubleton as a space of event based on a simple search engine. We employ set theory for defining doubleton and generate some properties.
\end{abstract}

\section{Introduction}

A search engine is extensively important to help users to find relevant information in Web. The search engines have different features, among of them are to service the tasks and subtasks that directly or indirectly uses the techniques such as indexing, filters, hub, page rank, hits, and etc \cite{croft2010}, but to access any information in Web the users need search term and other literal text in a query. In this case, the query has become the leading paradigm to find the information, whereby the information retreival (IR) is concerned with answering information need as accurately as possible. However, the difficult formulating of a query is always with the lack of understanding to special cases about the important information. The objective of this paper, therefore, to generate some adaptive properties of doubleton as semantic meaning of relation between a query and a search engine. 

\section{Related Works and Motivation}
In literal text, a name means persons and personas (including pseudonyms), organizations, corporate, and government bodies and families, or any entity such as "Social Network" or like the literal text of "Superficial Method for Extracting Social Networks for Academics using Web Snippets" \cite{nasution2011a}. Any literal text or name, other case we called it as term, consists of words or tokens, a word $w$ is the basic unit of discrete data, defined to be an item from a vocabulary indexed by $\{1,\dots,K\}$, 
\[
w_k = \cases{1 & if $k\in K$\cr
0 &  otherwise\cr}
\]
We defined some instances about a simple search engine \cite{nasution2012a}. 

\begin{definition}
\label{def:term}
A term $t_x$ consists of at least one or a set of words in a pattern, or $t_k = (w_1w_2\dots w_l)$, $l\leq k$, $k$ is a number of parameters representing word $w$, $l$ is the number of tokens (vocabularies) in $t_k$, $|t_k| = k$ is size of $t_k$. \kotak
\end{definition}

\begin{definition}
\label{def:searchengine}
Let a set of web pages indexed by search engine be $\Omega$, i.e., a set contains ordered pair of the terms $t_{k_i}$ and the web pages $\omega_{k_j}$, or $(t_{k_i},w_{k_j})$, $i=1,\dots,I$, $j = 1,\dots,J$. The relation table that consists of two columns $t_k$ and $\omega_k$ is a representation of $(t_{k_i},\omega_{k_j})$ where $\Omega_k = \{(t_k,\omega_k)_{ij}\}\subset\Omega$ or $\Omega_k = \{\omega_{k_1},\dots,\omega_{k_j}\}$. The cardinality of $\Omega$ is denoted by $|\Omega|$. \kotak
\end{definition}

\begin{definition}
\label{def:singleton}
Let $t_x$ is a search term, and $t_x\in{\cal S}$ where ${\cal S}$ is a set of singleton search term of search engine. A vector space $\Omega_x\subseteq\Omega$ is a singleton search engine event (\emph{singleton space of event}) of web pages that contain an occurrence of $t_x\in\omega_x$. The cardinality of $\Omega_x$ is denoted by $|\Omega_x|$. \kotak
\end{definition}

\begin{lemma}
\label{lem:singleton}
Let $t_x$ and $t_y$ are search term. If $t_x\ne t_y$, $t_x\cap t_y\ne\emptyset$ and $|t_y|<|t_x|$, then singleton search engine event of $t_x$ and $t_y$ is $\Omega_x = \Omega_x\cup\Omega_y$ or 
\begin{equation}
\label{pers:lemma1}
|\Omega_x| = |\Omega_x|+|\Omega_y|,
\end{equation} 
where $\Omega_x,\Omega_y\subseteq\Omega$. \kotakt
\end{lemma}

\begin{lemma}
\label{lem:counter}
If $t_y\ne t_z$ and $t_y\cap t_z = \emptyset$, then $|\Omega_y\cap\Omega_z|=0$ and $|\Omega_y\cup\Omega_z| = |\Omega_y|+|\Omega_z|$. \kotakt
\end{lemma}

\begin{lemma}
\label{lem:sama}
Let $t_x$ and $t_z$ are search terms. If $t_x\ne t_z$, $t_x\cap t_z=\emptyset$, and $\omega_x\cap\omega_z\ne\emptyset$, then $|\Omega_x|=|\Omega_z|$, $\Omega_x,\Omega_z\subseteq\Omega$. \kotakt
\end{lemma}

One singleton space of event is not same to another if their search terms are not same. Two singleton spaces of event have a distance or a similarity. 
Let $A$ be a set of search terms. A function $s : A\times A\rightarrow [0,1]$ is called \emph{similarity} (\emph{proximity}) on $A$ if $s$ is non-negative, symmetric, and if $s(t_x,t_y)\leq s(t_x,t_x)$ holds for all $t_x,t_y\in A$, with equality if and only if $\Omega_x =\Omega_y$ \cite{nasution2010, nasution2011c}.

In the context of modal logic, the similarity of two event spaces $\Omega_x$ and $\Omega_y$ for $\omega\Rightarrow t_x$ (true) and $\omega\Rightarrow t_y$ (true) respectively we use to explore the semantic relation of $\Omega$ where each search term is represented by a set of features \cite{nasution2012}. Let two different search terms $t_x\ne t_y$ for representing a same entities, $\Omega_x$ be most similar to $\Omega_y$ where $t_x$ is true, then $t_x\Rightarrow t_y$ will be true at $\Omega_y$ if and only if $t_y$ is true at $\Omega_x$, that is $\Omega_y(t_x)=1$ if $t_x$ is true at $\Omega_y$ then we have
\begin{equation}
\label{pers:simxy}
\Omega_y(t_x\Rightarrow t_y) = \Omega_x(t_y)
\end{equation}
where $\Omega_x(t_y)=1$ if $t_y$ is true at $\Omega_x$. Similarly, by symmetry on a similarity, we obtain also
\begin{equation}
\label{pers:simyx}
\Omega_x(t_y\Rightarrow t_x) = \Omega_y(t_x).
\end{equation}
and to generate a similarity of two singleton spaces of event, the singletons associate with a doubleton space of event. We define a doubleton space of event as follows.
\begin{definition}
\label{def:doubleton}
Let $t_x$ and $t_y$ are two different search term, $t_x\ne t_y$, $t_x,t_y\in{\cal S}$, where ${\cal S}$ is a set of singleton search term of search engine. A doubleton search term is ${\cal D} = \{\{t_x,t_y\}: t_x,t_y\in\Sigma\}$ and its vector space denoted by $\Omega_x\cap\Omega_y$ is a double search engine event (\emph{doubleton space of event}) of web pages that contain a co-occurrence of $t_x$ and $t_y$ such that $t_x,t_y\in\omega_x$ and $t_x,t_y\in\omega_y$, where $\Omega_x, \Omega_y, \Omega_x\cap\Omega_y\subseteq\Omega$.\kotak
\end{definition}

For example, one of widely the used measures for generating the relations in a social network extraction is Jaccard coefficient \cite{nasution2011b}, by using the singleton space and double space of events we have
\begin{equation}
\label{pers:jc}
s_{jc}(t_x,t_y) = \frac{|\Omega_x\cap\Omega_y|}{|\Omega_x|+|\Omega_y|-|\Omega_x\cap\Omega_y|}
\end{equation}
Similar to singleton space of event for $t_x$ and $t_y$ in a query, we have 
\begin{equation}
\label{pers:independent2}
\begin{array}{rcl}
\Omega_x\cap\Omega_y &=& (\Omega_x(t_x)=1)\wedge(\Omega_y(t_y)=1)\cr
                     &=& (\Omega_x(t_y)=1)\wedge(\Omega_y(t_x)=1)\cr
                     &=& (\Omega_x(t_x,t_y)=1) =(\Omega_y(t_x,t_y)=1)\cr
                     &\supseteq& \emptyset,\cr
\end{array}
\end{equation}
thus 
\begin{equation}
|\Omega_x\cap\Omega_y| = \sum_\Omega(\Omega_x(t_x,t_y)=1) = \sum_\Omega(\Omega_y(t_x,t_y)=1).
\end{equation}
So we obtain
\begin{equation}
|\Omega_{x_p}\cap\Omega_{y_p}| = \sum_\Omega(\omega_{x,y}\Rightarrow t_x,t_y) \leq |\Omega_x\cap\Omega_y|
\end{equation}
Therefore, the automatic extraction of meaning from the Web through using $\Omega$ affected by problem of singleton, i.e. a consequence in doubleton space of event.

\begin{problem}
\label{prob:doubleton}
Let $t_x$ and $t_y$ are two different search terms. If $t_x\in\Omega_x\cap\Omega_y$, then $t_x\in\Omega_x\cap\Omega_x$ and $t_x\in\Omega_y\cap\Omega_y$ such that
\begin{equation}
\label{pers:probdoubleton}
|\Omega_x\cap\Omega_y|\stackrel{?}{=}|\Omega_x\cap\Omega_y|+|\Omega_x\cap\Omega_x|+|\Omega_y\cap\Omega_y|
\end{equation}
\end{problem}

\section{The Adaptive Properties of Doubleton in Search Engine}

This Lemma \ref{lem:sama} explains that problem of singleton be $|\Omega_x|=|\Omega_y|$ if and only if $t_x\ne t_y$ but $t_x,t_y\in\omega_x$ $\wedge$ $t_x,t_y\in\omega_y$. In other word, based on combining equations,  $\Omega_x = \{(t_x,\omega_x)\} = \{(t_x,\omega_x\cup\omega_y)\} = \{(t_x,\omega_x)\cup(t_x,\omega_y)\} = \{(t_y,\omega_x)\cup(t_y,\omega_y)\} = \{(t_y,\omega_x\cup\omega_y)\}= \{(t_y,\omega_y)\} = \Omega_y$. This shows that the search terms may be different but they come from same web pages, and in this case they take the same meaning from web. 

Based on Lemma \ref{lem:singleton}, $|\Omega_x\cap\Omega_y| = |\{(t_x,\omega_x)\}\cap\{(t_y,\omega_y)\}| = |\{(t_x\cap t_y,\omega_x\cap\omega_y)\}| = |\{(t_y,\omega_y)\}| = |\Omega_y|$ or 
\begin{equation}
\label{pers:doubleton1}
|\Omega_x\cap\Omega_y|=|\Omega_y|
\end{equation}
Because $|\Omega_y|<|\Omega_x|$, we have $|\Omega_x\cap\Omega_y|<|\Omega_x|$. However, by Lemma \ref{lem:counter}, $|\Omega_x\cap\Omega_y| =  |\{(t_x,\omega_x)\}\cap\{(t_y,\omega_y)\}| = |\{(t_x\cap t_y,\omega_x\cap\omega_y)\}| = \emptyset$. This means that
\begin{equation}
\label{pers:doubleton2}
|\Omega_x\cap\Omega_y|<|\Omega_x| \wedge |\Omega_x\cap\Omega_y|<|\Omega_y|.
\end{equation}
Based on Lemma \ref{lem:sama}, $|\Omega_x\cap\Omega_y| = |\{(t_x,\omega_x)\}\cap\{(t_y,\omega_y)\}| = |\{t_x\cap t_y,\omega_x\cap\omega_y)\}| = |\{(t_x,\omega_x)\}| = |\Omega_x|$ or 
\begin{equation}
\label{pers:doubleton3}
|\Omega_x\cap\Omega_y|=|\Omega_x|
\end{equation}
Therefore, Eqs. (\ref{pers:doubleton1}), (\ref{pers:doubleton2}) and (\ref{pers:doubleton3}) clearly give $|\Omega_x\cap\Omega_y|\leq|\Omega_x|\leq|\Omega|$ or $|\Omega_x\cap\Omega_y|\leq|\Omega_y|\leq|\Omega|$, and this has proved the following theorem.

\begin{theorem}
\label{theorem:doubleton}
Let $t_x$ and $t_y$ are search terms. If $t_x\ne t_y$, but $\{(t_x,\omega_x)\}\cap\{(t_y,\omega)\}\ne\emptyset$, then a doubleton search engine event of $t_x$ and $t_y$ is the $\Omega_x\cap\Omega_y$, $\Omega_x,\Omega_y\subseteq\Omega$, $|\Omega_x\cap\Omega_y|\leq|\Omega_x|\leq|\Omega|$ and $|\Omega_x\cap\Omega_y|\leq|\Omega_y|\leq|\Omega|$. \kotakt
\end{theorem}

Disagree with it, let $t_x$ and $t_y$ are any search terms and we can derive a formula, that is, it starts from Eq. (\ref{pers:doubleton3}),  
\[
\begin{array}{rcll}
|\Omega_x\cap\Omega_y| &=& |\Omega_x| & \cr
                       &=& |\Omega_x|+|\Omega_y| & {\rm Lemma~\ref{lem:singleton}}\cr
&=& |\Omega_x|+|\Omega_x\cap\Omega_y| & {\rm Eq.~(\ref{pers:doubleton1})}\cr
&=& |\Omega_x|+|\Omega_y| + |\Omega_x\cap\Omega_y| & {\rm Lemma~\ref{lem:singleton}}\cr
\end{array}
\]
and we know that $|\Omega_x|=|\Omega_x\cap\Omega_x|$ and $|\Omega_y|=|\Omega_y\cap\Omega_y|$, then Eq. (\ref{pers:probdoubleton}) in Problem \ref{prob:doubleton} be
\begin{equation}
\label{pers:doubletonh}
|\Omega_x\cap\Omega_y| = |\Omega_x\cap\Omega_y|+|\Omega_x\cap\Omega_x|+|\Omega_y\cap\Omega_y|
\end{equation}
Thus Eq. (\ref{pers:doubletonh}) is a contraposition of Theorem \ref{theorem:doubleton}. In other word, to reduce the enumeration of singleton and in order to Eq. (\ref{pers:probdoubleton}) matches with Theorem \ref{theorem:doubleton}, there should $t_y\in\Omega_y$ satisfies $t_y\ne t_x$, but $t_x, t_y\in \Omega_x$.

\section{Conclusions and Future Work}
Studying to properties of relation between query and search engine gave the understanding about 
the semantic representation of doubleton for object in literal text. Our near future work is to generate some selective properties of search engine.

\end{document}